\documentclass[runningheads]{llncs}
\usepackage{setup}

\title{Controlling replication via the belief system in multi-unit organizations}
\titlerunning{Controlling replication via belief systems}

\author{Ravshanbek Khodzhimatov\inst{1}\orcidID{0000-0002-2761-2029} \and
Stephan Leitner \inst{2}\orcidID{0000-0001-6790-4651} \and
Friederike Wall\inst{2}\orcidID{0000-0001-8001-8558}}

\authorrunning{R. Khodzhimatov et al.}

\institute{Digital Age Research Center, University of Klagenfurt, 9020 Klagenfurt, Austria
\email{ravshanbek.khodzhimatov@aau.at}\\
\and
Department of Management Control and Strategic Management, University of Klagenfurt, 9020 Klagenfurt, Austria\\
\email{\{stephan.leitner, friederike.wall\}@aau.at}}

\begin{document}
\maketitle 
\begin{abstract}

Multi-unit organizations such as retail chains are interested in the diffusion of best practices throughout all divisions. However, the strict guidelines or incentive schemes may not always be effective in promoting the replication of a practice. In this paper we analyze how the individual belief systems, namely the desire of individuals to conform, may be used to spread knowledge between departments. We develop an agent-based simulation of an organization with different network structures between divisions through which the knowledge is shared, and observe the resulting synchrony. We find that the effect of network structures on the diffusion of knowledge depends on the interdependencies between divisions, and that peer-to-peer exchange of information is more effective in reaching synchrony than unilateral sharing of knowledge from one division. Moreover, we find that centralized network structures lead to lower performance in organizations.


\end{abstract}
\section{Introduction}
\label{sec:intro}

Multi-unit organizations are (potentially geographically) dispersed organizations that consist of a large number of divisions such as retail chain stores or fast-food franchises \cite{garvin08}. The divisions in multi-unit organizations predominantly operate in the same industry and promise customers the same brand experience in all divisions \cite{kim99}. To achieve this, the organizations have to make sure that divisions comply with the standards and best practices, but at the same time are given enough freedom to discover successful practices in the first place \cite{argote00,winter01}. 

To address this tension, organizations may employ different control mechanisms. Simons' Levers of Control framework \cite{simons95} identifies four levers that constitute a management control system. Diagnostic control systems are formal mechanisms that ensure that the branches work towards the agreed-upon goal (e.g., incentive schemes). Interactive control systems are formal information systems which give a focused view on the aspects of performance (e.g., KPIs). Boundary systems delineate the acceptable behavior in the organization (e.g., codes of conduct, franchise operations manuals) and can be enabling or constraining, depending on the management's decisions. Belief systems is a set of core organizational values and definitions that management uses to foster a desired environment (e.g., mission statements, organizational culture).

Belief systems in this context can be used to describe \textit{conformity}, which is defined as the internal desire of individuals to alter their behavior to match that of their peers. In contrast to compliance to organizational requirements, individuals conform voluntarily in pursuit of goals to blend into a team, gain approval of others, or increase accuracy of their actions by adopting the best practices of their peers \cite{cialdini04}. The actual desire to conform changes with cultural and demographic characteristics of individuals and with the environment and norms \cite{cialdini90}. However, the effect of desire of branch managers to conform on the actual adoption of the peers' practices depends on many factors, including the similarity in faced tasks \cite{argote95}, geographic proximity between branches \cite{galbraith90}, communication channels between branches \cite{hansen02}, and the rotation of employees \cite{kane05}. 

Chang and Harrington \cite{harrington00} studied the extent of centralization in retail chains in which managers come up with ideas for new practices. They found that organizations should employ boundary control systems that allow branches to experiment with new practices and routines subject to the constraint that branches need to adopt predetermined practices, and to combine this with diagnostic control systems by rewarding branch managers for replicating and passing along ideas. Garvin and Levesque \cite{garvin08} proposed diagnostic control systems that allow managers to prioritize their branch performance over the adoption of set practices. These studies implicitly assume that the best practices can be \textit{codified} and put in a guideline, ready for replication. However, Haldin-Herrgard \cite{haldin00} showed that this is usually not possible, and suggested decentralized (interpersonal) knowledge transfer mechanisms. Additionally, Garvin and Levesque \cite{garvin08} found that due to the large number of branches, it is difficult to enforce a centralized control in multi-unit organizations. In this context, the less studied lever of control, belief systems, may be more effective because they allow organizations to foster an environment for imitation without strict centralized control mechanisms \cite{tessier12}. 

In this study we are interested to what extent does the individuals' desire to conform affect the diffusion of knowledge between divisions, and how do different network structures through which the agents communicate affect this relation. We employ an agent-based simulations approach \cite{wall20,leitner15} to model the multi-unit organization and the $N\!K\!C\!S$-framework \cite{kauffman89,kauffman91} to model the environments in which the units (divisions) operate. The rest of the paper is structured as follows: Sec. \ref{sec:model} presents the method, Sec. \ref{sec:results} summarizes our findings, Sec. \ref{sec:conclusion} concludes the paper.

\section{Model}
\label{sec:model}
In this section we introduce the agent-based model of an organization with $P=5$ units. The task environment is based on the $N\!K\!C\!S$-framework \cite{kauffman89,levinthal97}. Agents make decisions to (a) increase their performance and (b) conform to the observed behavior of others. Sec. \ref{sec:design} introduces the task environment, Secs. \ref{sec:social} and \ref{sec:prefs} characterize the agents and describe how conformity is modeled. Sec. \ref{sec:discovering} describes the agents' search process, and Sec. \ref{sec:process} provides an overview of the sequence of events in the simulation.

\subsection{Task environment}
\label{sec:design}

We model an organization with $P=5$ agents (units), each of which faces a complex decision problem that is expressed as the vector of $N=4$ binary choices:
\begin{equation}
	\mathbf{x} = (
	\underbrace{x_1,x_2,x_3,x_4}_{\mathbf{x}^1},
	\dots, 
	\underbrace{x_{17},x_{18},x_{19},x_{20}}_{\mathbf{x}^5}
	),
	\label{eq:tasks}
\end{equation}
where bits $x_i \in \{ 0, 1 \}$ represent single tasks.
Every decision on a task $x_i$ yields a uniformly distributed performance contribution $\phi(x_i) \sim U(0,1)$. The decision problem is \textit{complex} in that the performance contribution $\phi(x_i)$, might be affected not only by the decision $x_i$, but also by decisions $x_j$, where $j \neq i $.

\begin{figure}[!tb]
	\centering
	\begin{minipage}{0.49\linewidth}
		\centering
		\includegraphics[width=\linewidth]{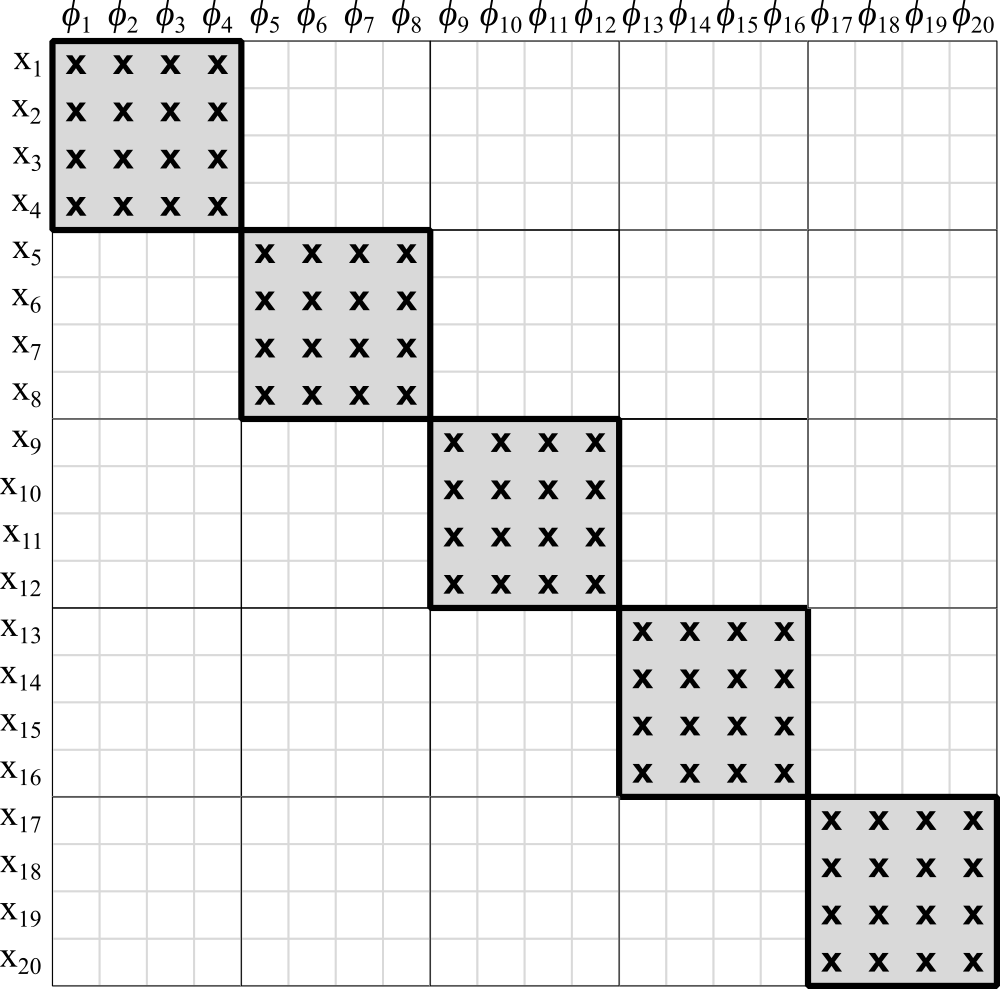}
		\caption*{(a) Internal interdependence}
	\end{minipage}
	\hfill
	\begin{minipage}{0.49\linewidth}
		\centering
		\includegraphics[width=\linewidth]{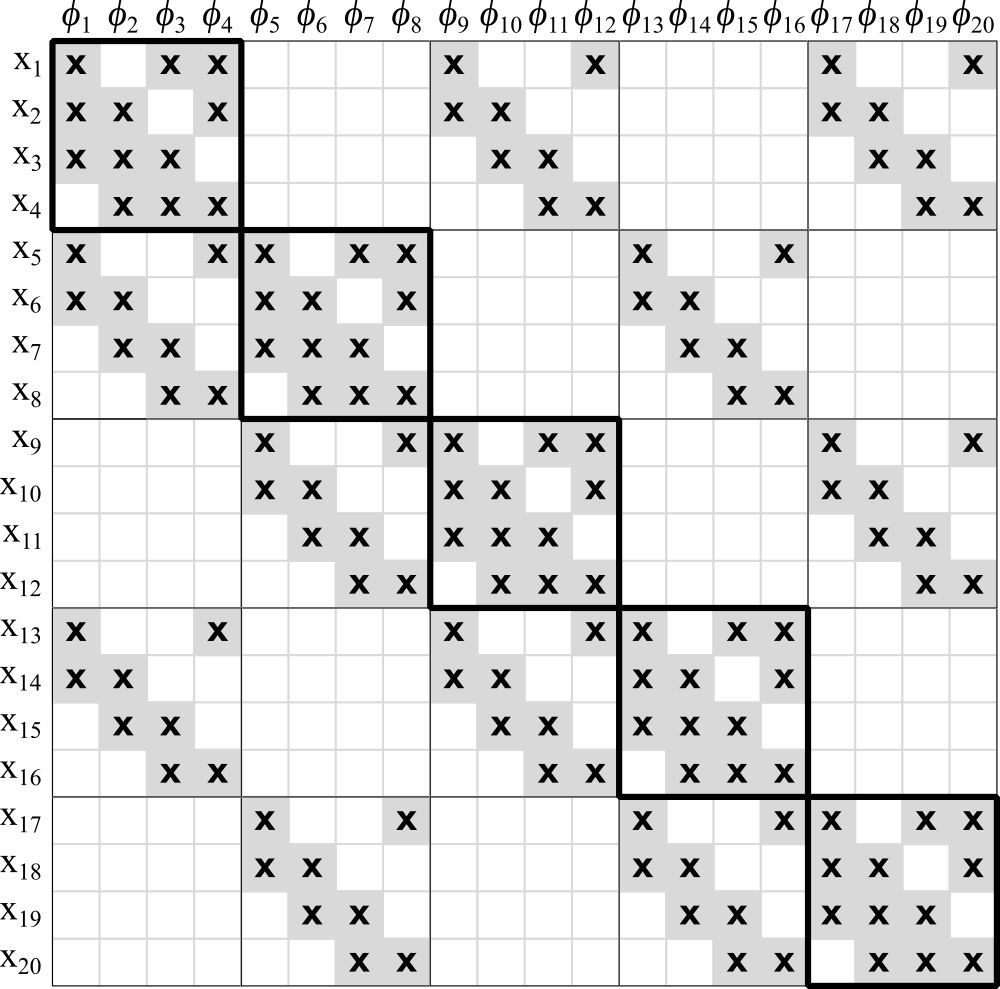}
		\caption*{(b) External interdependence}
	\end{minipage}
	
	\caption{Stylized interdependence structures with $P=5$ agents facing $N=4$ binary tasks. The crossed cells indicate inter-dependencies as follows: let $(i,j)$ be coordinates of a crossed cell in row-column order, then performance contribution $\phi(x_{i})$ depends on decision $x_j$.}
	\label{fig:interactions}
\end{figure}

We differentiate between two types of such interdependencies: (a) \textit{internal}, in which interdependence exists between the tasks within unit $p$, and (b) \textit{external}, in which interdependence exists between the tasks in units $p$ and $q$ for $p \neq q$. We control interdependencies by parameters $K,C,S$, so that every task interacts with exactly $K$ other tasks internally and $C$ tasks assigned to $S$ other agents externally \cite{kauffman91}:
\begin{equation}
	\phi (x_i) = \phi (x_i ,
	\underbrace{x_{i_1},...,x_{i_K}}_{\substack{K\text{ internal}\\\text{interdependencies}}}, \underbrace{x_{i_{K+1}} ,...,x_{i_{K + C \cdot S}}}_{\substack{C \cdot S \text{ external}\\ \text{interdependencies}}}),
	\label{eq:payoff}
\end{equation}
where $i_1, \dots , i_{K+C\cdot S}$ are distinct and not equal to $i$. The exact choice of the coupled tasks is random with one condition: every task affects and is affected by exactly $K+C\cdot S$ other tasks. In our analysis we consider two benchmark cases: (i) only internal interdependence ($K = 3, C = S = 0$) and (ii) internal and external interdependence: ($K=C=S=2$), as depicted in Fig. \ref{fig:interactions}.

Using Eq. \ref{eq:payoff}, we generate \textit{performance landscapes} as follows: for every task $x_i$ we generate performance contribution values corresponding to every combination of interdependent decisions from a uniform distribution. This results in a $N \times 2^{1+K+C \cdot S}$ matrix of uniform random numbers. We generate entire landscapes at the beginning of every simulation run to find the overall global maximum and normalize our results accordingly, to ensure comparability among different simulation runs.


At each time period $t$, agent $p$'s performance is a mean of performance contributions of tasks assigned to that agent:
\begin{equation}
	\phi^p (\mathbf{x}^p_t) = \frac{1}{N} \sum_{x_i \in \mathbf{x}^p_t} \phi(x_i),
	\label{eq:performance-agent}
\end{equation}
and the organization's overall performance is a mean of all agents' performances:
\begin{equation}
	\Phi (\mathbf{x}_{t}) = \frac{1}{P} \sum_{p=1}^{P} \phi^p(\mathbf{x}^p_t)~.
	\label{eq:performance-org}
\end{equation}

All agents in the multi-unit organization operate in (not perfectly) similar environments, i.e., the same decisions of agents tend to lead to the same performance, with minor differences stemming from their local environments. We model this similarity between task environments of different units using the pairwise correlations between their performance landscapes \cite{verel13}:
\begin{equation}
	\label{eq:corr}
	\textbf{corr} \left( \phi^p(\mathbf{x}^p_i), \phi^q(\mathbf{x}^q_i) \right) = \rho \in [0,1],
\end{equation}
for all $1 \leq i \leq N$ and $p \neq q$. In our analysis we use the value of $\rho=0.9$ as our benchmark, as it represents high similarity in units with a few local differences.

\subsection{Conformity metric}
\label{sec:social}

\begin{figure}
	\centering
	\begin{minipage}{0.25\linewidth}
		\includegraphics[width=\linewidth]{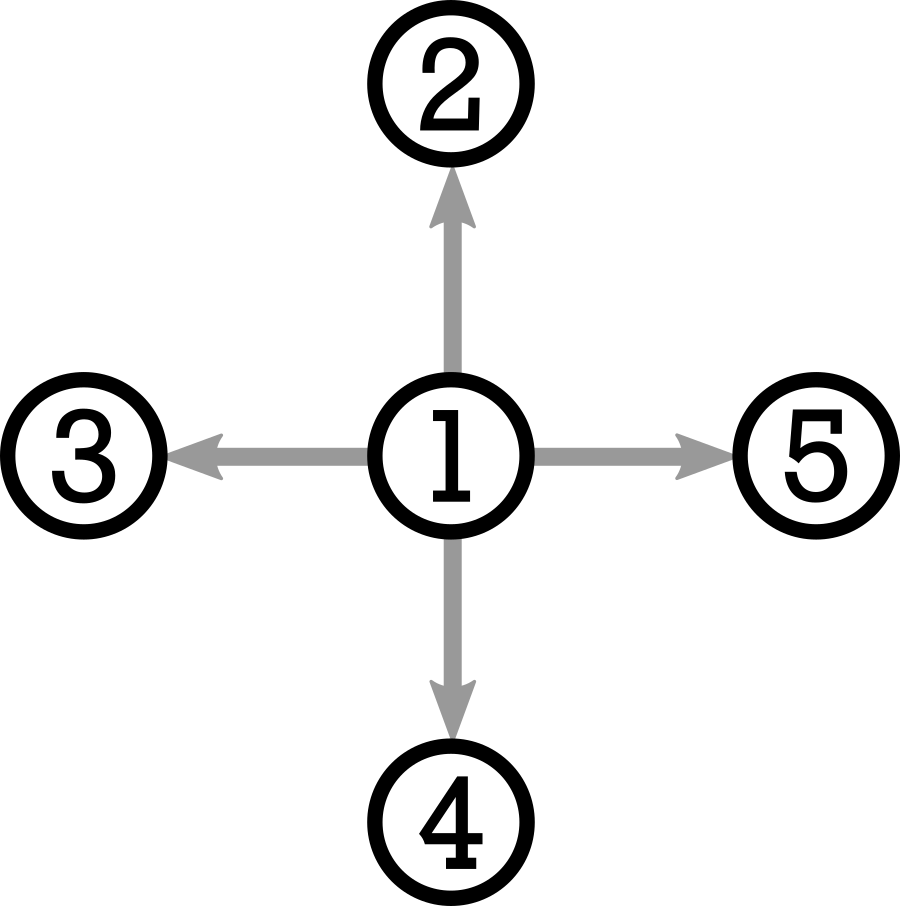}
		\caption*{(a) Star network}
	\end{minipage}
	\hfill
	\begin{minipage}{0.25\linewidth}
		\includegraphics[width=\linewidth]{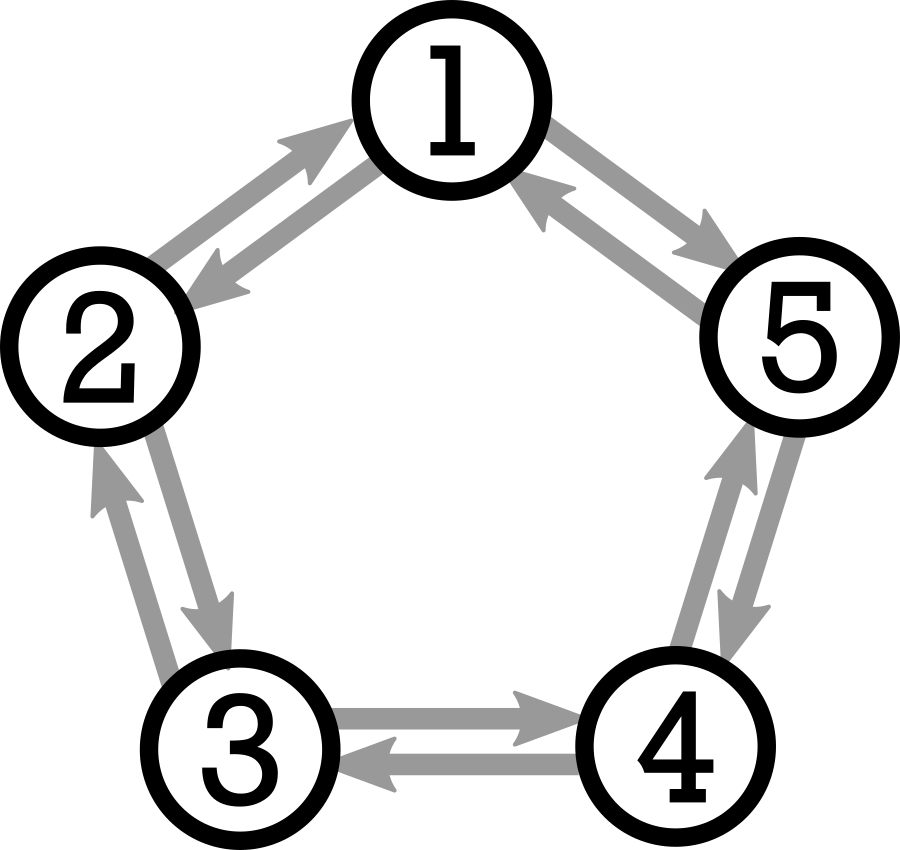}
		\caption*{(b) Ring network}		
	\end{minipage}		
	\hfill
	\begin{minipage}{0.25\linewidth}
		\includegraphics[width=\linewidth]{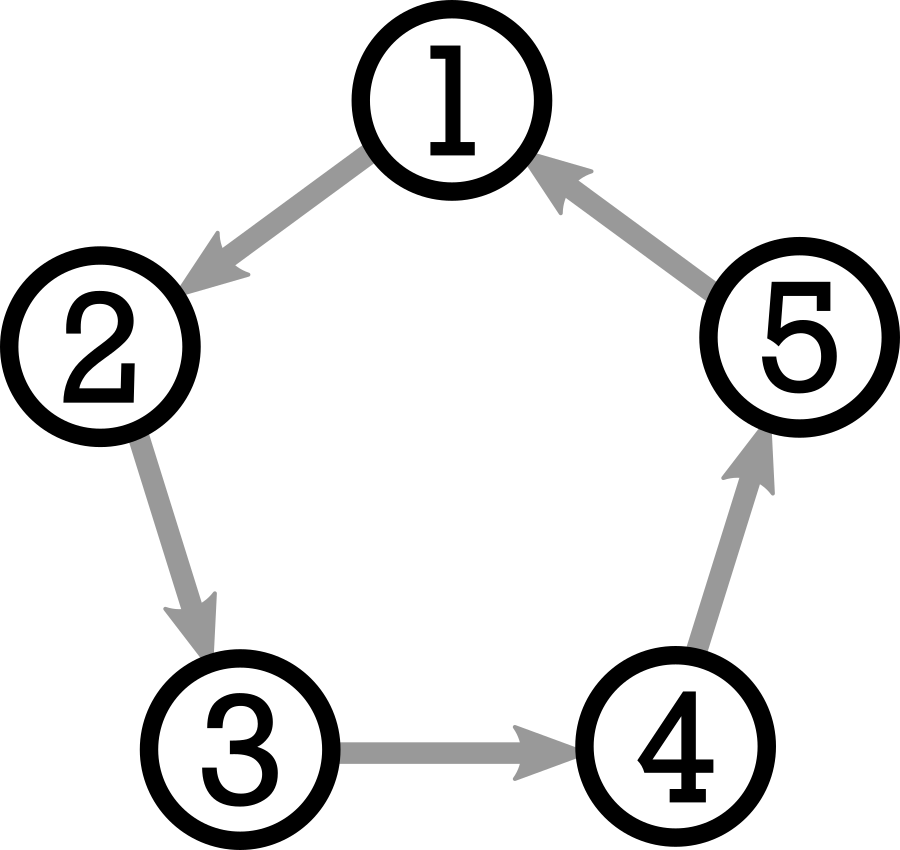}
		\caption*{(d) Cycle network}
	\end{minipage}

	\vspace{2em}

	\begin{minipage}{0.5\linewidth}
		\includegraphics[width=\linewidth]{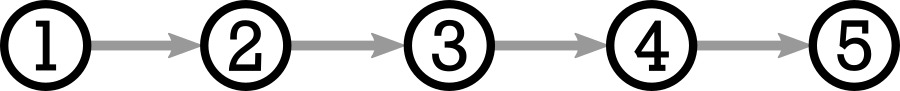}
		\caption*{(e) Line network}		
	\end{minipage}
	\caption{Network structures in which $P=5$ agents (nodes) share information through directed links. Star and Line networks capture unidirectional transfer of knowledge and Ring and Cycle networks capture mutual transfer of knowledge.}
	\label{fig:network}
\end{figure}

To measure conformity we implement our version of the Social Cognitive Optimization algorithm introduced by Xie et al. \cite{xie02}. At every time step $t$, agents share the decisions they have made on their tasks with the fellow agents, according to one of the network structures described in Fig. \ref{fig:network}, where nodes represent agents and the directed links represent sharing of information.\footnote{Including network structures in the model is a major extension over the papers using a similar approach to model the spread of information in organizations \cite{hojimat21a,hojimat21b}.} Every agent $p$ stores the shared information in the memory set $L^p$ for up to $T_L=50$ periods, after which the information is ``forgotten'' (removed from $L^p$).

The measure of conformity of agent $p$'s decisions $\mathbf{x}^{p}_{t}$ is computed as the average of the matching bits in the memory:
\begin{equation}
	\phi_{conf} (\mathbf{x}^p_{t}) = 
	\begin{cases}
		\displaystyle\frac{1}{\lvert L^p_t \rvert \cdot N } \sum_{\mathbf{x^L} \in L^p_t} \sum^{N}_{i=1} [x^p_i==x^L_i], & t > T_L \\
		0, & t \leq T_L
	\end{cases}
	\label{eq:norms}
\end{equation}
where $\lvert L^p_t \rvert$ is the number of entries in agent $p$'s memory at time $t$, and the statement inside the square brackets is equal to $1$ if true, and $0$ if false \cite{iversion62}.

\subsection{Agents' preferences}
\label{sec:prefs}

We model the agents' preferences as a linear function \cite{akerlof80,gali94,tversky91} of performance $\phi^p$ and conformity metric $\phi_{conf}$ (see Eqs. \ref{eq:performance-agent} and \ref{eq:norms}):

\begin{equation}
	u^p(\mathbf{x}^p) = \alpha \cdot \phi^p (\mathbf{x}^p) + \beta \cdot \phi_{conf} (\mathbf{x}^p)
	\label{eq:utility}
\end{equation}

where $\alpha + \beta = 1$.

\subsection{Search process}
\label{sec:discovering}

In line with Simon \cite{simon57}, our agents are \textit{boundedly rational}. In particular, the agents are not global optimizers and want to increase their utility given limited information: at time $t$, agents can observe their own performance in the last period, $\phi^p (\mathbf{x}^p_{t-1})$, and the decisions of all team members in the last period \textit{after} they are implemented, $\mathbf{x}_{t-1}$.

In order to come up with new solutions to their decision problems, agents perform a search in the neighbourhood of $\mathbf{x}_{t-1}$ as follows: agent $p$ randomly switches one decision $x_i \in \mathbf{x}^p$ (from $0$ to $1$, or vice versa), and assumes that other agents will not switch their decisions (Levinthal \cite{levinthal97} describes situations in which agents switch more than one decision at a time as \textit{long jumps} and states that such scenarios are less likely to occur, as it is hard or risky to change multiple processes simultaneously). We denote this vector with one switched element by $\hat{\mathbf{x}}^p_t$. 

Next, the agent has to make a decision whether to stick with the status quo,  $\mathbf{x}^p_t$, or to switch to the newly discovered $\hat{\mathbf{x}}^p_t$. The rule for this decision is to maximize the utility function defined in Eq. \ref{eq:utility}:

\begin{equation}
	\mathbf{x}^p_t = \underset{\mathbf{x} \in \{\mathbf{x}^p_{t-1}, \hat{\mathbf{x}}^p_{t}\}}{\arg\max} u(\mathbf{x}) ,
	\label{eq:goalprog}
\end{equation}

\subsection{Process overview, scheduling and main parameters}
\label{sec:process}
\begin{figure}[tb]
	\centering
	\includegraphics[width=\linewidth]{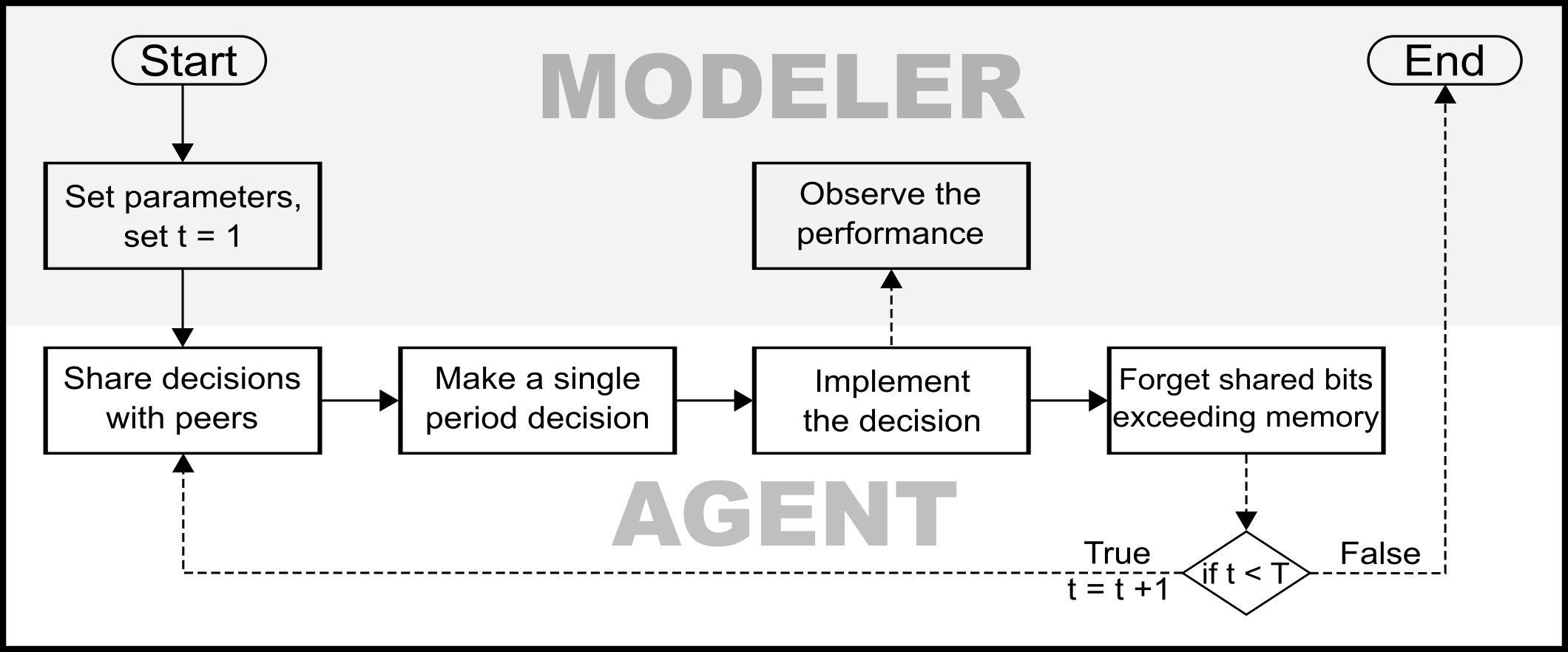}
	\caption{Process overview}
	\label{fig:funcflowchart}
\end{figure}

The simulation model has been implemented in \textit{Python 3.8} and \textit{Numba} just-in-time compiler. Every simulation round starts with the initialization of the agents' performance landscapes, the assignment of tasks to $P=5$ agents. For reliable results, we generate the entire landscapes before the simulation, which is feasible for $P=5$ given modern computing limitations, and the initialization of an $M=20$ dimensional bitstring as a starting point of the simulation run. After initialization, agents perform the \textit{hill climbing} search procedure outlined above and share information regarding their own decisions according to the network structure. The observation period $T$, the memory span of the employees $T_L$, and the number of repetitions in a simulation, $R$, are exogenous parameters, whereby the latter is fixed on the basis of the coefficient of variation. Fig. \ref{fig:funcflowchart} provides an overview of this process and Tab. \ref{tab:params} summarizes the main parameters used in this paper.

\begin{table}[tb]
	\centering
	\begin{tabular}{clc}
		Parameter & 
		\multicolumn{1}{c}{Description}				                & Value \\
		\hline
		$M$       & Total number of tasks                       & 20 \\
		$P$       & Number of agents                            & 5  \\
		$N$       & Number of tasks assigned to a single agent  & 4  \\
		$[K,C,S]$ & Internal and external couplings             & $[3,0,0]$, $[2,2,2]$\\                  
		$\rho$    & Pairwise correlation between landscapes     & 0.9 \\
		$T_L$     & Memory span of agents                       & 50 \\
		$T$       & Observation period                          & 500 \\
		$R$       & Number of simulation runs per scenario      & $1,\!000$ \\
		$[\alpha, \beta]$ & Weights for performance $\phi$ and conformity $\phi_{conf}$  &  $[1,0]$, $[0.5,0.5]$, $[0,1]$\\
		\hline
	\end{tabular}
	\vspace{1em}
	\caption{Main parameters}
	\label{tab:params}
\end{table}

\section{Results}
\label{sec:results}

In this section we present selected findings from running $R=1,\!000$ simulations for 4 different network structures and task environments with and without external interdependencies between departments. In each simulation scenario we observe organization-level performance and the measure of synchrony across divisions for $T=500$ time periods.  Sec. \ref{sec:simmeasure} defines the synchrony measure, and  Secs. \ref{sec:similarity} and \ref{sec:performance} present the findings.

\subsection{Measure of synchrony}
\label{sec:simmeasure}

To measure the synchrony of a strategy we first define the Hamming distance, which is a metric that returns the number of bits that are distinct in two bit strings. For example, the Hamming distance between a bit string $1001$ and $1101$ is equal to $1$, as they differ in only one bit, and flipping just one bit is sufficient to make them equal. Similarly, the Hamming distance between identical bit strings $1001$ and $1001$ is equal to zero.

Next, we define the \textit{asynchrony} or distinctness of a bitstring as the sum of all pairwise Hamming distances between the bit sub-strings allocated to different agents.

\begin{equation}
	\mathbf{H}(\mathbf{x}) = \sum_{p=1}^{P} \sum_{q=p}^{P} H(\mathbf{x}^p, \mathbf{x}^q)
\end{equation}

Finally, we measure the synchrony of a bitstring as a complement of the asynchrony normalized by its maximum:

\begin{equation}
	\mathbf{S}(\mathbf{x}) = 1 - \frac{\mathbf{H(\mathbf{x})}}{\max \{ \mathbf{H} (\mathbf{x})\}}
\end{equation}

\subsection{Findings regarding synchrony}
\label{sec:similarity}

\begin{figure}
	\centering
	\includegraphics[width=0.5\linewidth]{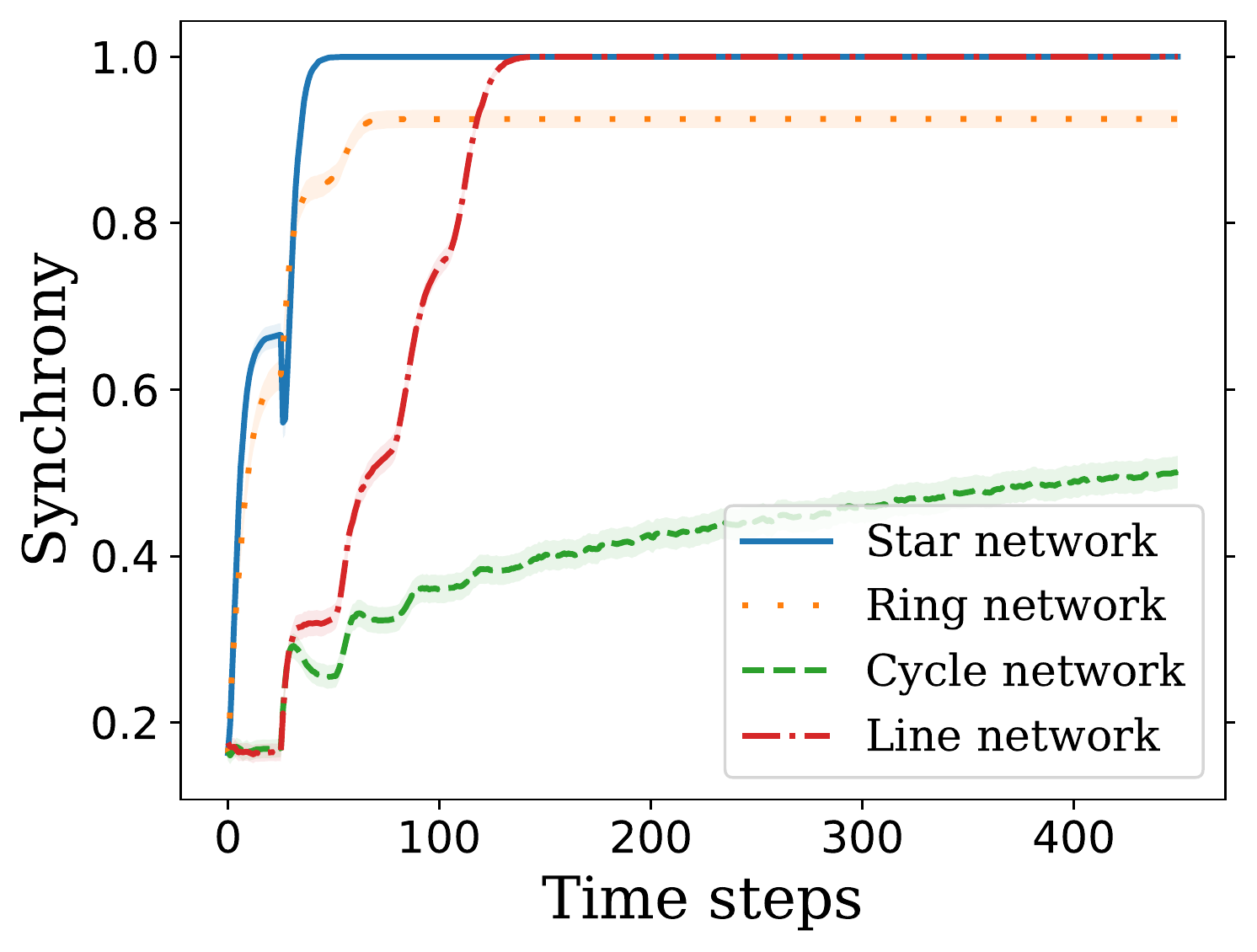}
	\caption{Synchrony measure for full conformity ($\alpha=0$, $\beta=1$). This figure applies to all scenarios with $P=5$, and, by construction, is not affected by the structure of the task environment.}
	\label{fig:fullconf}
\end{figure}

In this section we analyze how the synchrony in the organization is affected by the individuals' desire to conform by the different network structures through which they communicate. solely prefer the conformity and do not consider the actual performance in their departments (i.e., $\alpha=0$, $\beta=1$) to understand how the network structures operate to spread the best practices between departments. Fig. \ref{fig:fullconf} shows that the top-to-bottom unilateral sharing of knowledge in Star and Line network structures leads to the full synchrony, with Line network being slower to converge. The peer-to-peer egalitarian network structures like Cycle network lead only to partial synchrony even if the agents have a full desire to conform. By construction, this relation holds for all task environments with $P=5$ agents.

This intuitive finding, however, no longer holds in a more realistic scenario, in which agents have both performance and conformity in their preferences (i.e. $\alpha=\beta=0.5$). Indeed, we find that centralized Star network leads to a high synchrony only in the short term in the absence of external interdependencies between divisions. In the long term, however, the Cycle network leads to a higher synchrony. In presence of external interdependencies between divisions, the Ring network leads to the highest synchrony and passes the Star network after 50 periods. Moreover, the Line network leads to the lowest synchrony for interrelated divisions. All of these scenarios, however, lead to a significantly higher synchrony than the situations in which agents do not have a desire to conform (i.e., $\alpha=1$, $\beta=0$), which lead to a synchrony measure $\mathbf{S}(\mathbf{x}) \leq 0.3$.

\begin{figure}
	\centering
	\begin{minipage}{0.48\linewidth}
		\includegraphics[width=\linewidth]{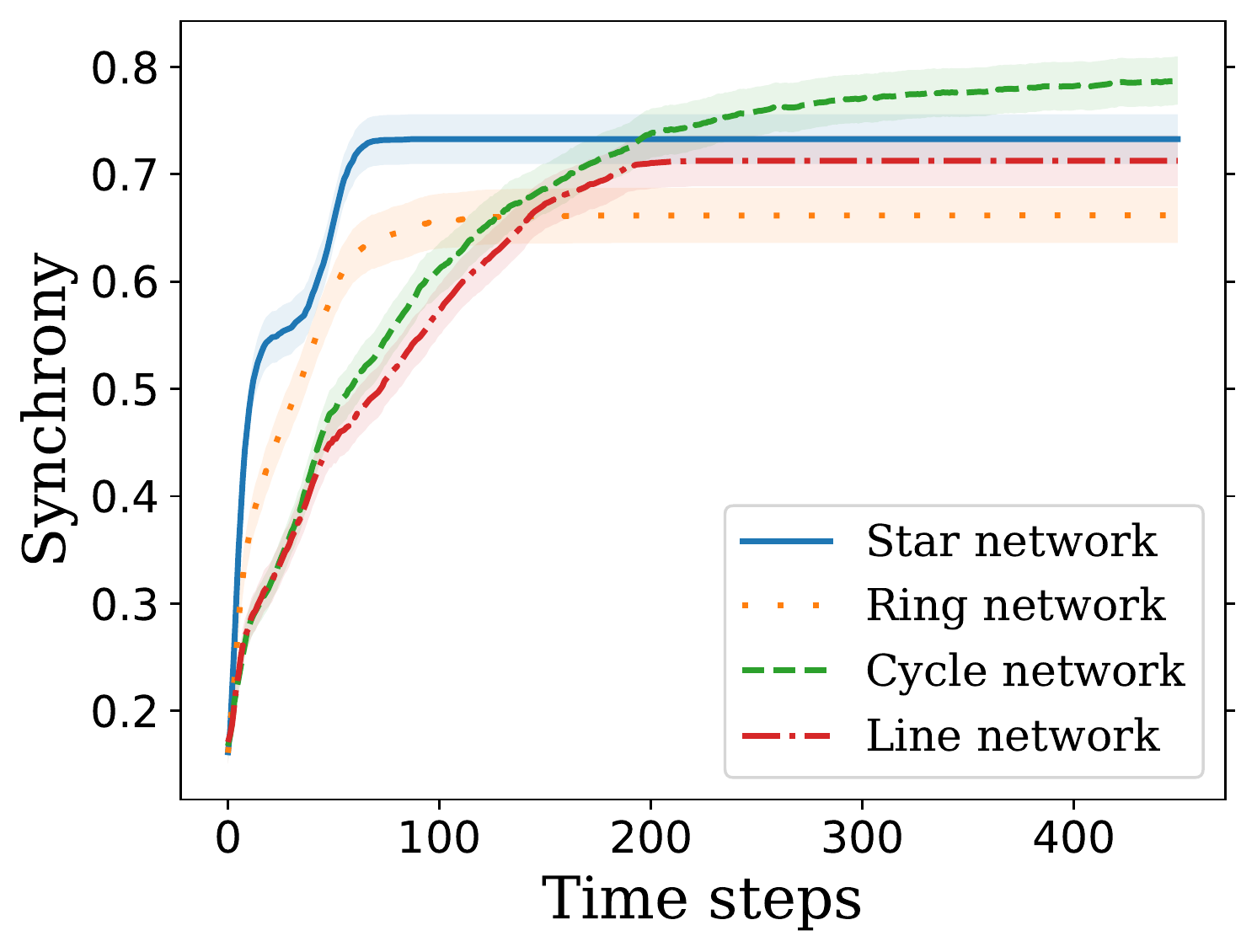}
		\caption*{(a) Internal interdependence}
	\end{minipage}
	\begin{minipage}{0.48\linewidth}
		\includegraphics[width=\linewidth]{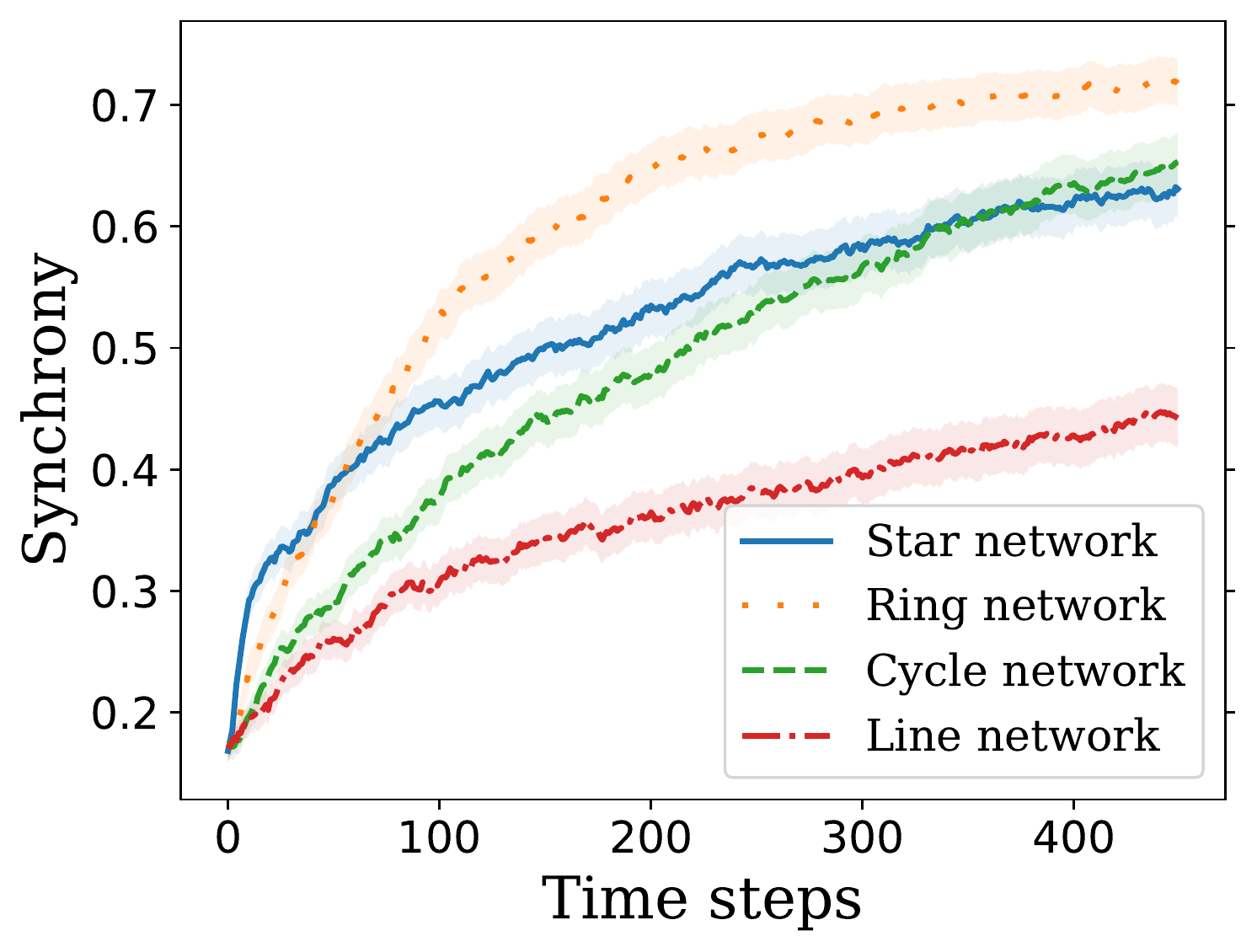}
		\caption*{(b) External interdependence}
	\end{minipage}
	\caption{Synchrony measure for moderate level of conformity}
\end{figure}

These results indicate that, while conformity can significantly increase the diffusion of best practices in the organizations, the management should be careful in promoting it and consider the nature of tasks the units are facing and the interdependence between them. We find that the naive idea of identifying the successful unit and promoting other units to directly replicate its practices does not always lead to the highest diffusion of knowledge, and that forsaking a centralized control and promoting a peer-to-peer communication is more effective in the long run.  

\subsection{Performance measure}
\label{sec:performance}

Next, we look at how the different network structures affect the organization-level performance for environments with and without external interdependencies. We find that the centralized Line and Star network structures actually lead to less organizational performance than the decentralized Ring and Cycle networks. This happens because in the centralized network structures, the central units do not see decisions of their peers and, thus, cannot benefit from the knowledge they have gained. In the decentralized network structures, on the other hand, all units directly or indirectly observe the decisions made by their peers, and, thus, can benefit from the knowledge of all departments.

Between the latter two, the Ring network leads to the higher performance in presence of external interdependencies between units, and the Cycle network leads to the highest long-term performance in the absence of them.

\begin{figure}
	\centering
	\begin{minipage}{0.48\linewidth}
		\includegraphics[width=\linewidth]{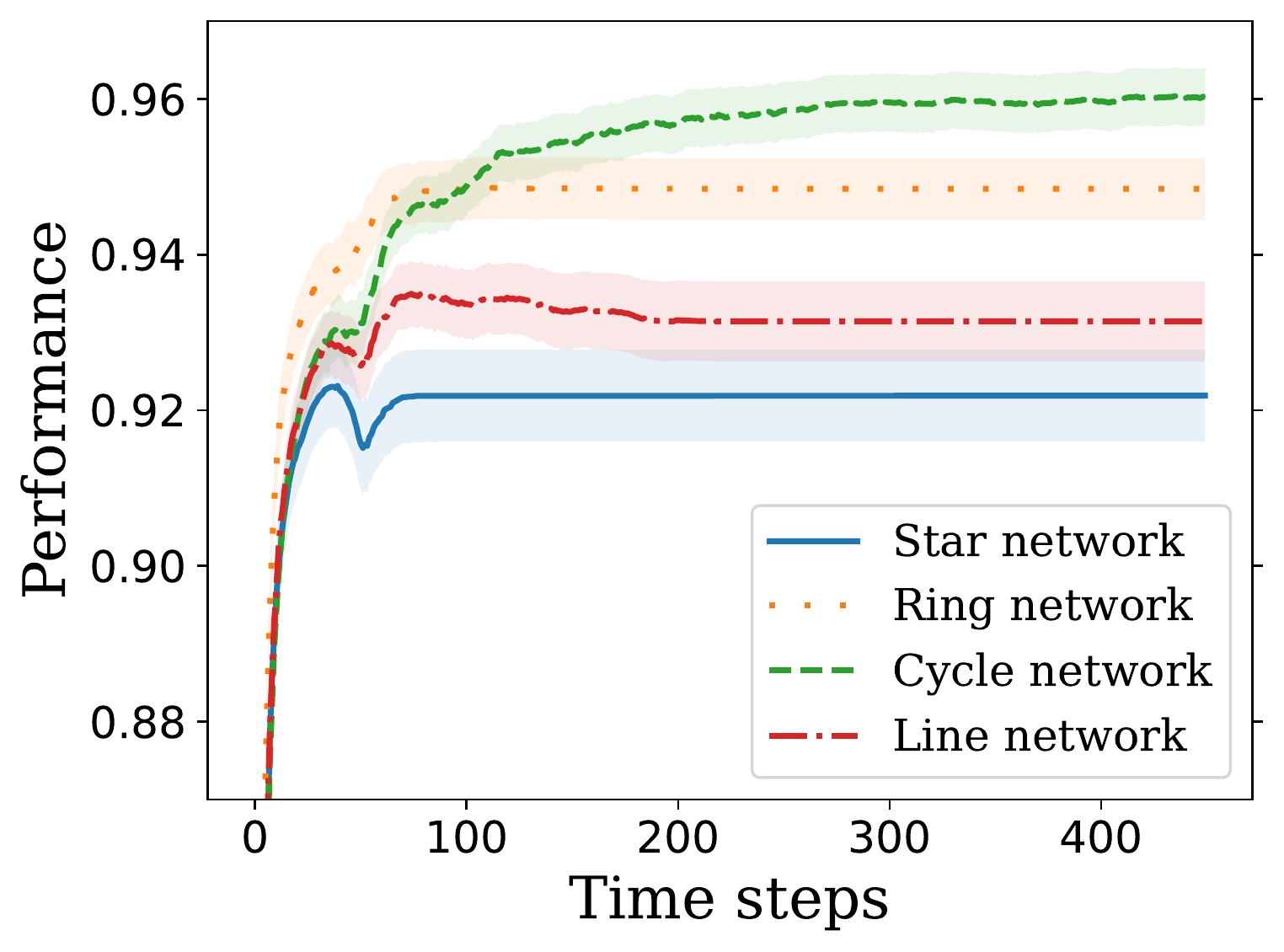}
		\caption*{(a) Internal interdependence}
	\end{minipage}
	\begin{minipage}{0.48\linewidth}
		\includegraphics[width=\linewidth]{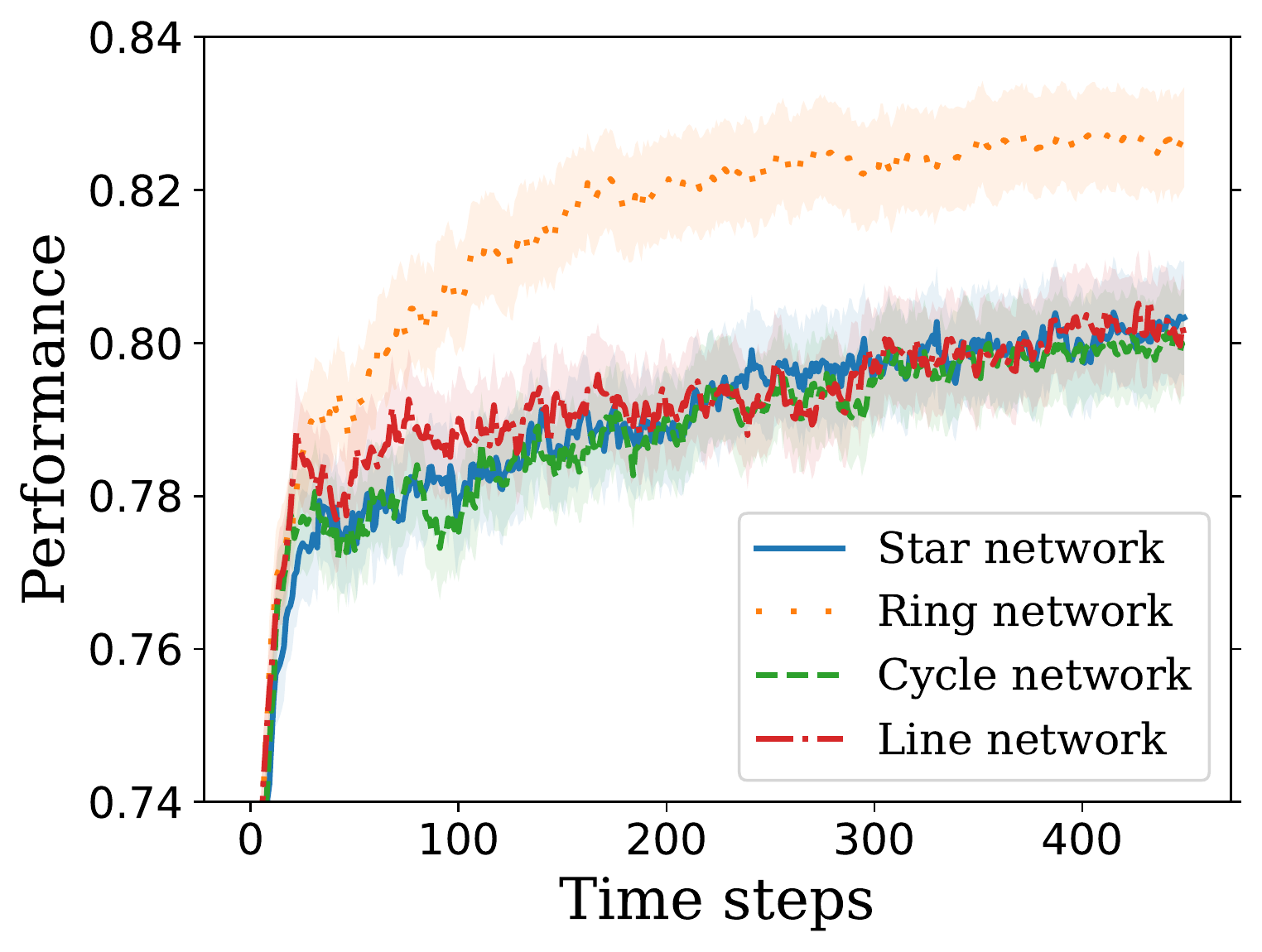}
		\caption*{(b) External interdependence}
	\end{minipage}
	\caption{Performance for moderate level of conformity}
\end{figure}
\section{Conclusion}
\label{sec:conclusion}

In this paper we studied how multi-unit organizations can employ individuals' belief systems, particularly, their desire to conform to the behavior of their peers, to achieve the diffusion of best practices between their units. We performed an agent-based simulation of the organization and compared the achieved synchrony for different network structures between the units. We found that, contrary to the intuition, the centralized spread of knowledge from a single unit to others leads to a lower long-term synchrony than the decentralized peer-to-peer sharing of knowledge between all units. Interestingly, our results do not feature a trade-off between organizational performance and synchrony -- we find that in most situations the decentralized Cycle and Ring networks help to achieve both the high synchrony and the high performance.

The implications of our results are that the management in multi-unit organizations should forsake the centralized control over the diffusion of knowledge (via diagnostic or boundary control systems) and to promote an organizational culture of sharing knowledge and conforming to the most frequent practices. The limitations of our research include the lack of historical performance in the agents' consideration to conform -- further research might address this via dynamically updated weights for the desire to conform. 

\bibliographystyle{splncs04}
\bibliography{refs.bib}
\end{document}